\documentclass[10pt,conference]{IEEEtran}
\IEEEoverridecommandlockouts
\usepackage{cite}
\usepackage{amsmath,amssymb,amsfonts}
\usepackage{algorithmic}
\usepackage{graphicx}
\usepackage{textcomp}
\usepackage{xcolor}
\usepackage{xspace}
\usepackage{fontawesome}
\usepackage{multirow}
\usepackage{rotating}
\usepackage{tikz}
\usepackage{soul}
\usepackage{makecell}
\usepackage{url}
\usepackage{enumitem}
\usepackage{booktabs}
\usepackage{fontawesome}
\usepackage{pifont}
\usepackage{amssymb}
\usepackage[colorlinks=true, allcolors=black]{hyperref}
\usepackage{tcolorbox}
\usepackage[framemethod=TikZ]{mdframed}
\usepackage{arydshln}

\newcommand{\ie}{\emph{i.e.,}\xspace}
\newcommand{\eg}{\emph{e.g.,}\xspace}
\newcommand{\etc}{etc.\xspace}
\newcommand{\etal}{\emph{et~al.}\xspace}
\newcommand{\secref}[1]{Section~\ref{#1}\xspace}

\newcommand{\figref}[1]{Fig.~\ref{#1}\xspace}

\newcommand{\tabref}[1]{Table~\ref{#1}\xspace}

\newcommand\REVISED[1]{#1}

\newboolean{showcomments}
\setboolean{showcomments}{true}

\ifthenelse{\boolean{showcomments}}
  {\newcommand{\nb}[2]{
    \fbox{\bfseries\sffamily\scriptsize#1}
    {\sf\small$\blacktriangleright$\textit{#2}$\blacktriangleleft$}
   }
  }
  {\newcommand{\nb}[2]{}
  }

\newcommand{\invited}{82\xspace}
\newcommand{\answered}{40\xspace}
\newcommand{\accepted}{40\xspace}
\newcommand{\participants}{32\xspace}
\newcommand{\participantsSelected}{29\xspace}

\newcommand{\recordingtime}{50 hours\xspace}

\newcommand{\assignedReviews}{120\xspace}
\newcommand{\submittedReviews}{96\xspace}
\newcommand{\selectedReviews}{72\xspace}

\newcommand{\program}[1]{{\small{\texttt{#1}}}}

\def\BibTeX{{\rm B\kern-.05em{\sc i\kern-.025em b}\kern-.08em
    T\kern-.1667em\lower.7ex\hbox{E}\kern-.125emX}}

\begin{document}

\title{Deep Learning-based Code Reviews:\\A Paradigm Shift or a Double-Edged Sword?}

\author{
\IEEEauthorblockN{
Rosalia Tufano\textsuperscript{\textdagger*}, 
Alberto Martin-Lopez\textsuperscript{\textdagger*}, 
Ahmad Tayeb\textsuperscript{\textdaggerdbl}, 
Ozren Dabi\'c\textsuperscript{\textdagger}, 
Sonia Haiduc\textsuperscript{\textdaggerdbl}, 
Gabriele Bavota\textsuperscript{\textdagger}}
\IEEEauthorblockA{
\textsuperscript{\textdagger}\textit{Software Institute -- USI Universit\`{a} della Svizzera italiana, Switzerland}\\\textsuperscript{\textdaggerdbl}\textit{Florida State University, United States}\\
\textsuperscript{\textdagger}\{rosalia.tufano, alberto.martin, ozren.dabic, gabriele.bavota\}@usi.ch, 
\textsuperscript{\textdaggerdbl}\{atayeb2, shaiduc\}@fsu.edu}
}

\maketitle

\begingroup\renewcommand\thefootnote{*}
\footnotetext{Equal contributor. Author order determined by coin flip.}
\endgroup

\begin{abstract}
Several techniques have been proposed to (partially) automate code review. Early support consisted in recommending the most suited reviewer for a given change or in prioritizing the review tasks. With the advent of deep learning in software engineering, the level of automation has been pushed to new heights, with approaches able to provide feedback on source code in natural language as a human reviewer would do. Also, recent work documented open source projects adopting Large Language Models (LLMs) as co-reviewers. Although the research in this field is very active, little is known about the actual impact of including automatically generated code reviews in the code review process. While there are many aspects worth investigating (\eg is knowledge transfer between developers affected?), in this work we focus on three of them: (i) review quality, \ie the reviewer's ability to identify issues in the code; (ii) review cost, \ie the time spent reviewing the code; and (iii) reviewer's confidence, \ie how confident is the reviewer about the provided feedback. We run a controlled experiment with \participantsSelected professional developers who reviewed different programs with/without the support of an automatically generated code review. During the experiment we monitored the reviewers' activities, for over \recordingtime of recorded code reviews. We show that reviewers consider valid most of the issues automatically identified by the LLM and that the availability of an automated review as a starting point strongly influences their behavior: Reviewers tend to focus on the code locations indicated by the LLM rather than searching for additional issues in other parts of the code. The reviewers who started from an automated review identified a higher number of \emph{low-severity} issues while, however, not identifying more \emph{high-severity} issues as compared to a completely manual process. Finally, the automated support did not result in saved time and did not increase the reviewers' confidence.
\end{abstract}

\begin{IEEEkeywords}
Code review, Controlled Experiment
\end{IEEEkeywords}

\section{Introduction} \label{sec:intro}

Code review is an essential activity in both industrial and open source projects. Its benefits have long been recognized and studied and include improved code quality and a reduced incidence of bugs, among others \cite{sadowski:icse2018,bacchelli:icse2013,morales:saner2015,bavota:icsme2015,mcintosh:msr2014}. However, code review can also be time consuming and costly \cite{bosu:esem2013}, and over the past decade, researchers have been studying ways to reduce the cost of this activity while maintaining its benefits. Some early efforts included recommending the most suitable reviewer for a given change \cite{balachandran:icse2013,thongtanunam:saner2015,xia:icsme2015,ouni:icsme2016}, predicting the defectiveness of a patch before or after being reviewed \cite{soltanifar:esem2016,sharma:spe2019}, and classifying the usefulness of review comments \cite{pangsakulyanont:iwesep2014,rahman:msr2017}.
\eject

Over the last few years, along with the rise of Deep Learning (DL), came a new wave of approaches aimed at reducing the costs of code review by exploiting DL techniques to automate the code review process. These approaches have been able to provide natural language review comments about source code similar to what a software developer would provide \cite{li:fse2022,tufano:icse2022,li.l:esecfse2022}. Moreover, recent work reported open source projects adopting Large Language Models (LLMs) as co-reviewers \cite{Tufano:msr2024}. 

Given the tremendous potential shown by LLMs in helping developers with various software engineering tasks \cite{Liu:icse2024,Ma:icse2024,Junjielong:icse2024,Nam:icse2024}, it seems natural to involve them also in the code review process, with the goal of reducing developer effort and software cost. However, little is known about the actual impact of including automatically generated code reviews in the code review process. For example, it is hard to anticipate how using automated code reviews as a starting point could impact the knowledge transfer between developers or the quality of the final code review, or if using automated reviews could lead to biases or blind spots in developers' analysis of the code, if it leads to a lower code review cost, \etc Studying the impact of all these factors goes beyond the scope of a single paper. However, we aim to make the first steps in this direction by focusing on three specific aspects, namely: (i)~code review quality, \ie a reviewer's ability to identify issues in the code; (ii) code review cost, \ie the time spent by developers reviewing the code; and (iii) reviewer's confidence, \ie how confident the reviewer is about the provided feedback. 

To determine the impact that using automated code reviews can have on these three aspects, we present a controlled experiment involving \participantsSelected developers. The participants performed a total of \selectedReviews reviews across six projects written either in Python or Java in which we injected quality issues representative of problems usually found during code review \cite{mantyla2008types}. Participants have been assigned to one of the two languages based on their expertise. Each review was performed with one of three treatments. The first, named \REVISED{\emph{manual code review} (MCR)}, assumes no availability of an automatically generated review as starting point, thus reflecting the classic code review scenario. The second, named \emph{automated code review} (ACR), provides reviewers with a review automatically generated by ChatGPT Plus~\cite{chatgpt}, which they can use as a starting point for their final reviews (\eg they could discard generated comments, rephrase them, or complement the set of identified issues). 

\eject

This scenario is representative of the current state of the art in automated code review \cite{Tufano:msr2024}. The third, named \emph{comprehensive code review} (CCR), aims at simulating a hypothetical scenario in which tools for the automated generation of code reviews reached a whole new level where they can correctly identify all the important issues in the code. To simulate this scenario, we provide participants with a code review correctly pointing to all issues we injected, presenting it as automatically generated. The latter scenario, while not realistic nowadays given the current technology, allows us to observe the extent to which reviewers would trust an automated tool by adopting its suggestions and the impact it would have on reviewing time. 
Our main findings can be summarized as follows:

\begin{enumerate}

\item \emph{Reviewers considered valid most of the issues identified by ChatGPT Plus}. On average, 89\% of the issues automatically identified by the LLM have been kept by the reviewers in their final review.


\item \emph{The availability of an automated review as a starting point strongly influences the reviewer’s behavior}. Reviewers mostly focused on the code locations pointed out in the automatically generated review they were provided with (this holds for both ACR and CCR treatments). While we observed substantial variability in the code locations commented on by reviewers who inspected a program manually, the ones who started from an automated review tended to focus on the code locations already provided in it. 

\item \emph{The automated code review generated by the LLM does not help in identifying more high-severity issues as compared to a completely manual process.} We only observed a significant difference in the number of \emph{low-severity} issues in the code (in favor of the treatment adopting the ChatGPT-based review). 

\item \emph{Even assuming an excellent support in terms of automated review (CCR treatment), reviewers do not save time with respect to a fully manual process}. This is due to the fact that they need to interpret the automatically generated comments and check for their correctness in the code.

\item \emph{Providing reviewers with automatically generated reviews (both in the ACR and CCR treatment) does not impact their confidence, which is comparable to that observed in the \REVISED{MCR} treatment}. This might be due to the fact that the provided review does not help in better understanding the program, but only highlights potential issues.

\end{enumerate}

Study material and data are publicly available \cite{replication}.
\section{Study Design} \label{sec:design}

The \emph{goal} of the study is to assess the impact that having access to automatically generated code reviews has on the code review process. More specifically, we aim to understand how the availability of an automated code review affects the quality of the final review written by the reviewer, the time they spend reviewing the code, and their confidence about the submitted review. \REVISED{Note that our study focuses on a scenario in which the output of the DL-based approach is provided to the reviewer as a support for the review writing. 

However, these tools could also be seen as a possible support for the contributor, \ie a first feedback loop \emph{before} submitting the code for the actual (human) review (see \eg \cite{Vijayvergiya:google2024}).
While this work focuses on the perspective of reviewers and the bias they may be subject to when using automated code reviews, future work could certainly investigate the contributors' perspective (\ie do automated reviews affect the quality of the code reviewed by humans?).
}

The study addresses the following research questions (RQ):

\textbf{RQ$_0$}: \emph{Is there a statistically significant difference in the characteristics of the code reviews written by developers with/without automated support?}
This preliminary RQ aims at quantitatively comparing the code reviews submitted by developers with/without access to automatically generated reviews. We analyze various aspects including the number of issues reported, the length of the code review (number of sentences), and the problematic code locations identified (\eg number of different lines in which issues have been found).

\textbf{RQ$_1$}: \emph{To what extent does having access to automated code reviews increase the likelihood of identifying code quality issues?}
We inject quality issues in the code of our software projects and assess the extent to which participants were able to identify them with/without the support of automated code review. Since we manually inspect all \selectedReviews code reviews written by developers, we also analyze, discuss, and compare among the different treatments the additional quality issues that were not injected by us, but were found by participants.

\textbf{RQ$_2$}: \emph{To what extent does the availability of automated code reviews save time during the review process?}
We compare the amount of time spent by reviewers with/without the support of automated code review. In particular, we analyze the time spent: (i)~to complete the review task, \ie overall time; (ii)~inspecting the code; and (iii)~writing the actual review.

\textbf{RQ$_3$}: \emph{Does the availability of automated code reviews increase the reviewers' confidence?}
At the end of each reviewing task, we ask reviewers to rate their confidence in the submitted review, and we investigate if the availability of automated code reviews has an impact on the reviewers' confidence.

\subsection{Context Selection}
\label{sec:context}


\subsubsection{Participants}
\label{sec:participants}

We used convenience sampling \cite{sedgwick:bmj2013,Stratton:prehospital2021} to recruit \participantsSelected participants who (i) are currently professional developers (28 of them) or (ii) have worked in the past as professional developers and are now enrolled in a CS graduate program (one of them). We did not include any CS student without at least one year of industrial experience. Although this limited the number of participants we could involve, we consider industrial experience essential in any study on code review, given its common use in industrial practice. For simplicity, in the following we use the term ``developers'' when referring to participants, even though one of them is not \emph{currently} working as a developer.
We invited \invited developers to participate in our study, asking each of them to contribute three code reviews, each using a different treatment (details in \secref{sec:treatments}). 

The invitation email, available in our replication package \cite{replication}, asked them to accept the invitation if they (i) are familiar with code review; and (ii) have experience with at least one of the two subject programming languages (\ie Java and Python). Their availability was collected using a Google form. Only if they accepted our invitation, we asked them four questions. The first was: \emph{``Please select the programming languages in which you have expertise (check both of them if you are familiar with both)''}, with possible answers being Python and Java. The information collected was used to assign participants to the code review tasks described in \secref{sec:programs} (\ie to ensure that they were only allocated code review tasks involving a programming language they were familiar with). 

Then, the developers answered three questions related to their expertise: their years of experience in programming, their current role/position, and whether they took part in the past in the code review process as a reviewer, as a developer whose code was reviewed, in both roles, or in none of them. We received an answer from \answered developers, all of whom accepted to participate in the study. However, in the end, only \participants completed at least one of the tasks assigned (\ie at least one code review). From these, we selected \participantsSelected participants for our analyses, in such a way as to have the same number of participants per system and treatment (details in \secref{sec:analysis}).
\REVISED{On average, the \participantsSelected participants have 11.4 years of programming experience (median=10, min=3, max=35); three of them selected Java as programming language, six Python, and 20 checked both languages. Finally, three of them have not been involved in code review in the past (while still being familiar with it), one only as a reviewer, one only as a contributor, while 24 have covered both roles.}

\begin{table}[t]
  \caption{Summary of the object programs.\vspace{-0.2cm}}
  \label{tab:tasks}
  \centering
  \resizebox{\columnwidth}{!}{%
  \begin{tabular}{lllrr}
      \toprule
      \textbf{Project ID} & \textbf{Language} & \textbf{Source} & \textbf{LoC} & \textbf{Issues} \\
      \midrule
      \program{maze-generator} & Java & Rosetta~\cite{rosetta} & 164 & 1 \\
      \program{maze-generator} & Python & Rosetta~\cite{rosetta} & 75 & 2 \\
      \program{number-conversion} & Java & Rosetta~\cite{rosetta} & 116 & 4 \\
      \program{number-conversion} & Python & Rosetta~\cite{rosetta} & 81 & 2 \\
      \program{stopwatch} & Java & Apache~\cite{apache-commons-lang} & 528 & 7 \\
      \program{stopwatch} & Python & Translated & 258 & 4 \\
      \program{tic-tac-toe} & Java & Rosetta~\cite{rosetta} & 326 & 2 \\
      \program{tic-tac-toe} & Python & Rosetta~\cite{rosetta} & 121 & 7 \\
      \program{todo-list} & Java & Artificial & 206 & 3 \\
      \program{todo-list} & Python & Artificial & 198 & 3 \\
      \program{word-utils} & Java & Apache~\cite{apache-commons-lang} & 509 & 6 \\
      \program{word-utils} & Python & Translated & 426 & 7 \\
      \bottomrule
  \end{tabular}
  }
  \vspace{-0.4cm}
\end{table}

\subsubsection{Programs}
\label{sec:programs}
\tabref{tab:tasks} shows the programs that we asked participants to review. We considered six different projects, each available in both programming languages, \ie Java and Python. Our object programs are taken from different sources (see column ``Source'' in \tabref{tab:tasks}). We selected three programs (\program{maze-generator}, \program{number-conversion} and \program{tic-tac-toe}) from Rosetta Code~\cite{rosetta}, a repository of programming tasks written in multiple languages, including Java and Python. Two Java programs (\program{stopwatch} and \program{word-utils}) were taken from utility classes of the Apache Commons Lang library~\cite{apache-commons-lang} and then translated by the authors into Python. 

One program (\program{todo-list}) was created from scratch, for both Java and Python. To ensure that the implementation of the selected programs was of high quality, each of them was reviewed by two authors of the paper. Additionally, all programs were reviewed by a professional developer with seven years of experience and high familiarity with both Python and Java.
The selection of the programs was guided by two main goals: (i) to ensure code reviews are manageable in terms of both time and complexity for our study participants; we therefore chose programs that are relatively small in terms of lines of code (refer to the ``LoC'' column in \tabref{tab:tasks}); and (ii) to avoid requiring specific domain knowledge from the participants; for this reason we selected programs that any seasoned developer could easily understand and review. 

The chosen programs include: \program{maze-generator}, which creates a random maze in the console based on user-specified dimensions; \program{number-conversion}, enabling the conversion of decimal numbers into binary, octal, hexadecimal, and Roman numeral formats; \program{stopwatch}, a basic program that performs stopwatch functions like start, stop, reset, and split time; \program{tic-tac-toe}, an implementation of the corresponding game played via command line interface (CLI) against an agent programmed not to lose; \program{todo-list}, a CLI-based to-do list manager supporting adding, removing, prioritizing and listing tasks; and \program{word-utils}, a set of utility functions offering string manipulation utilities such as capitalization, case swapping, and abbreviation.

\begin{figure}
\centering
  \includegraphics[width=0.95\linewidth]{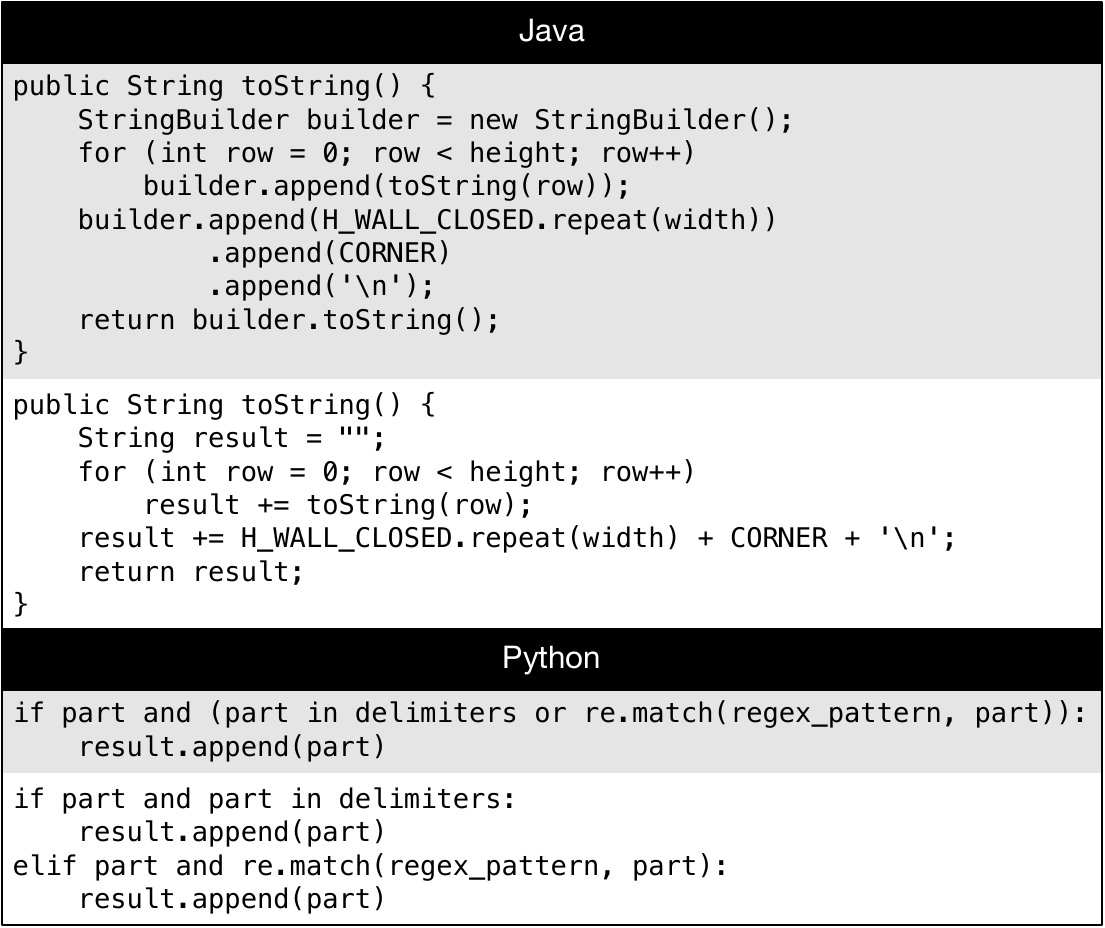}
  \vspace{-0.2cm}
  \caption{Examples of injected issues in Java and Python. The top part of each example represents the original code; the lower part is the code after injection.}
  \label{fig:injected_issue}
  \vspace{-0.4cm}  
\end{figure}

We then manually injected a number of quality issues in the selected programs given that, among other things, we aim at assessing the extent to which an automated code review increases the chances of identifying code quality issues. The column ``Issues'' in \tabref{tab:tasks} shows the number of issues injected into every program. Overall, we injected 48 issues across the 12 programs. The complete list of issues injected and their description is available in our replication package~\cite{replication} and includes code duplication, structural defects (\eg overly long methods), documentation issues (\eg mismatches with respect to the implementation) and logic bugs, among others. We paid attention not to inject issues which can be very easily detected by participants (\eg bugs that make the program crash), since they might not represent realistic simulations of code submitted by a developer for review. This also meant that we did not inject a fixed number of issues per project because we found a different number of issues to be suitable for different programs. The injected issues are inspired by the taxonomy of issues found in code reviews documented by M\"antyl\"a and Lassenius~\cite{mantyla2008types}. \REVISED{In particular, M\"antyl\"a and Lassenius found 77\% of the issues identified by reviewers to be related to evolvability defects (\eg documentation issues, sub-optimal implementation choices), with the remaining 23\% pertaining functional issues (\eg wrong implementation logic). By classifying the type of issues we injected according to the definitions given by Fregnan \etal \cite{fregnan:emse2022}, we injected: 78\% (64\%) evolvability issues and 22\% (36\%) defects in the object Java (Python) programs. 

\eject

As an additional note, the injected issues cover 73\% of the issue types in the taxonomy by M\"antyl\"a and Lassenius~\cite{mantyla2008types}. For example, we did not inject \emph{visual representation} issues, since code formatting usually depends on each project's practices.}
\figref{fig:injected_issue} shows two examples of injected issues (one per language). The top part of each example shows the original code, while the bottom part reflects the code after the issue injection. The Java example represents the injection of an issue related to performance: we replaced the use of \texttt{StringBuilder} with string concatenation, creating a degradation of performance. The Python code, on the other hand, exemplifies the injection of a structural defect: we introduced an unnecessary nested condition, also creating duplicated code. Each object program also featured: (i) a \texttt{main} file allowing to run it; and (ii) test cases exercising its basic functionalities.  

\subsection{Code Review Treatments}
\label{sec:treatments}

To understand the impact of supporting developers with code review automation, we defined three \emph{treatments}. The first, named \REVISED{\emph{manual code review} (MCR)}, resembles the classic code review process performed by developers without any automated support. The second, named \emph{automated code review} (ACR) provides the participant with a code review automatically generated by ChatGPT Plus (\ie GPT-4)~\cite{chatgpt}. The prompt used to create code reviews was \emph{``Provide a detailed code review of the following $<$Java/Python$>$ program: $<$code$>$''}. 
The third treatment, called \emph{comprehensive code review} (CCR), is designed to simulate the scenario where the reviewer is provided with an ``automated'' code review which correctly identifies all issues we injected. Unlike the first two treatments, which represent realistic scenarios in industry, this CCR treatment simulates an ideal, hypothetical scenario where the automated code review is able to identify all quality issues in a given code. 

While this scenario is not yet fully attainable with today's technology, studying it can provide valuable insights into how reviewers' behaviors might change if they had access to a (future) tool capable of identifying every quality issue in the code (\eg would they trust the tool enough to significantly save time?). To simulate this scenario, participants in the CCR treatment were told that the reviews were automatically generated, even though this was not the case. More specifically, we first manually performed the code review and made sure to capture all the issues we injected in the review. Then, we used ChatGPT Plus to rephrase the code reviews we wrote using the prompt: \emph{``Rephrase the following code review comment as if you are generating it: $<$comment$>$. The comment refers to the following $<$Java/Python$>$ code: $<$code$>$''}. Note that, while the CCR reviews include exactly $n$ comments identifying the $n$ issues injected in each program, the ACR treatment may identify comments that only address some (or none) of the injected issues. Additionally, ACR could include comments on other quality issues we did not introduce. 

The 24 code reviews generated or rephrased by ChatGPT Plus for ACR and CCR (6 projects $\times$ 2 languages $\times$ 2 treatments) are publicly available~\cite{replication}.

In summary, the three treatments compare the classic, manual code review process (\REVISED{MCR}) with the automation available in practice nowadays (ACR) and a Utopian scenario we hope to reach one day in the field (CCR). 




\subsection{Experimental Setup and Procedure}
\label{sec:procedure}

Our study is comprised of 36 different \emph{code review tasks} (6 projects $\times$ 2 languages $\times$ 3 treatments). As previously mentioned, we asked participants to review three programs, each with a different code review treatment. The three tasks were all in the same programming language but related to different projects. 
We provided each participant with instructions to connect, via the Remote Development plugin~\cite{remote-vscode} of Visual Studio Code (VS Code), to a server we set up with the environment needed to run the study. We took care of installing the versions of Java (17) and Python (3.10) needed to run any of the object programs. Also, we installed in VS Code the Java Extension Pack~\cite{java-vscode} and Python~\cite{python-vscode}. 
Once connected, the participants could see the review tasks assigned to them directly in the IDE without the need of installing/configuring anything. In particular, the participants were presented with three projects already imported in the IDE, two of which included a code review (for the treatments \emph{automated code review} and \emph{comprehensive code review}). The code review was presented through the Code Review plugin in VS Code~\cite{codereview-vscode}, which also allows to mark source files with review comments, and to modify/delete the already provided comments part of the provided code review.
All projects featured a \texttt{README} file with instructions on where to find a description of the program to review, how to use the Code Review plugin, how to run the program and associated tests, and a reminder to rate the confidence of their review at the end of each code review task, by simply writing a score from 1 (very low confidence) to 5 (very high confidence) at the end of the \texttt{README} file. 

This was the only action required from the participants at the end of each task. The Code Review plugin of VS Code~\cite{codereview-vscode} took care of storing the final version of their code review on our server, \ie the code review including all comments the participants manually wrote plus, for the ACR and CCR treatments, the comments they decided to keep (as is or by rephrasing/modifying them) from the provided reviews. Each comment is linked to a file and a range of selected text (in terms of line/column numbers).

Lastly, besides the final code review and the self-assessed confidence, we also monitored the participants' behavior in the IDE using Tako~\cite{tako-vscode,minelli:icpc2015}, a VS Code plugin collecting the actions performed in the IDE. Tako records events such as opening and closing files and tabs, switching between them and editing files, among others. This allows us to perform the time-based analyses needed to answer RQ$_2$.


\subsection{Data Analysis}
\label{sec:analysis}
Out of the \REVISED{\assignedReviews} reviews we assigned (\REVISED{\accepted} participants $\times$ 3 treatments), \REVISED{\submittedReviews} of them were completed by \REVISED{\participants} participants. 
As a result, we ended up with a different number of reviews for different programs for each treatment. In order to compare the treatments fairly, we systematically selected the highest possible number of reviews per treatment such that all treatments featured exactly the same number of reviews on exactly the same programs (thus being comparable). \REVISED{We prioritized reviews from participants who completed two (7 participants) or three (18) tasks.} This led to the selection of \REVISED{\selectedReviews} reviews (\ie \REVISED{24} per treatment) from \REVISED{\participantsSelected} participants. 

Two authors independently inspected each review to extract data needed to answer our RQs which cannot be automatically collected. At the end, they had a meeting to discuss about differences in the extracted data and agree on the correct data to report. The manually collected data include:

\begin{enumerate}
\item \emph{The total number of reported quality issues in each review}. This information cannot be automatically extracted by counting the number of comments in the code review, since each comment may point to several quality issues, even in the same code location. 

\item \emph{The number (and percentage) of injected issues identified in the code review}. 

\item \emph{The number (and percentage) of quality issues identified in the initial reviews provided to participants which have been kept in the finally submitted code review}. This metric has only been collected for code reviews resulting from the ACR and CCR treatments.

\item \emph{The additional quality issues (\ie unrelated to the injected ones) present in the final code review}. 
\end{enumerate}

Out of the \selectedReviews reviews manually inspected by two authors, they disagreed only on three reviews about the total number of reported quality issues, and about the number of injected issues identified. Conflicts were solved via open discussion. 


\begin{table}[h!]
	\centering
	\scriptsize
	\caption{\REVISED{Variables used in our study.}\vspace{-0.2cm}}
        \label{tab:variables}
  \resizebox{\columnwidth}{!}{
	\begin{tabular}{lm{5.1cm}}
		\toprule
		\textbf{Variable} & \textbf{Description}\\\midrule
		
		\multicolumn{2}{c}{\emph{Dependent Variables (RQ$_0$)}}\\\midrule
		Number of reported quality issues \hspace{-0.15cm} & The total number of quality issues reported in a submitted review\\\hdashline
		Length of the code review & The number of sentences in a submitted review\\\hdashline
		Covered code locations & Number of lines subject of at least one review comment\\\midrule
		
		\multicolumn{2}{c}{\emph{Dependent Variables (RQ$_1$)}}\\\midrule
		Is injected issue identified & The participant found the injected issue\\\midrule
		
		\multicolumn{2}{c}{\emph{Dependent Variables (RQ$_2$)}}\\\midrule
		Total time & The total time spent on the whole code review session\\\hdashline
		Time reviewing & The time spent on the actual code to review (\ie excluding the time spent writing the review, or running the program)\\\hdashline
		Time writing & The time spent writing review comments or reading the ones already provided in the automated review\\\midrule
		
		\multicolumn{2}{c}{\emph{Dependent Variables (RQ$_3$)}}\\\midrule
		Confidence & The confidence score provided by the participant for the submitted review\\\midrule
		
		\multicolumn{2}{c}{\emph{Independent Variables}}\\\midrule
		Treatment & The treatment used to perform the code review (one among MCR, ACR, CCR)\\\midrule
		
		\multicolumn{2}{c}{\emph{Control Variables (Experience)}}\\\midrule
		Years of experience & The years of experience in programming of the participant\\\hdashline
		Involved in code review & Whether the participant has been involved in the code review process as a reviewer, as a developer whose code was reviewed, in both roles, or in none of them.\\\midrule
		
		\multicolumn{2}{c}{\emph{Control Variables (Review task)}}\\\midrule
		Programming language & The language in which the code to be reviewed is written (Java or Python)\\\hdashline
		Program & The program on which the review task must be performed\\\hdashline
		Issue type & The type of the issue to spot, classified according to the work by Fregnan \etal \cite{fregnan:emse2022}\\
		\bottomrule
	\end{tabular}
  }
	\vspace{-0.6cm}
\end{table}

\REVISED{We answer our RQs by comparing the code reviews output of the three treatments. \tabref{tab:variables} reports: the dependent variables considered in each RQ, the independent variable, being the same for all RQs (\ie \emph{treatment}), and the control variables. 

For \textbf{RQ$_0$}, we compare: (i) the total number of reported quality issues; (ii) the length of the code review in terms of number of sentences; and (iii) the covered code locations, in terms of number of lines subject of at least one review comment (we also differentiate between code statements and code documentation). For answering \textbf{RQ$_1$}, we compare the ability of participants to identify injected issues during code review. In this case, a boolean dependent variable (``\emph{Is injected issue identified}'') has been used to indicate whether each of the bugs injected in the programs under study has been identified.}

\REVISED{Concerning \textbf{RQ$_2$}, we exploit the data collected by Tako~\cite{tako-vscode} to compare: (i) the total time spent on the whole code review session, (ii) the time spent on the actual code to review (\ie excluding the time spent writing the review, or running the program), and (iii) the time spent writing review comments or reading the ones already provided in the automated review. Finally, for \textbf{RQ$_3$} we compare the confidence scores provided by participants to check whether the availability of the automated code review had an impact on the perceived confidence.} 

\REVISED{For all comparisons we use boxplots to visualize the distributions. In addition to that, we run the following statistical analyses. When needed, we opted for non-parametric tests since all our distributions are not normally distributed (Shapiro-Wilk test). 

For \textbf{RQ$_1$}, we build a multivariate logistic regression model having \emph{Is injected issue identified} as the dependent variable and \emph{treatment} as independent variable with MCR set as the reference level to more easily look at the impact of introducing automation (\ie ACR and CCR treatments) in the code review process. We include all control variables listed in \tabref{tab:variables}.}

\REVISED{Concerning the other RQs (\ie \textbf{RQ$_0$}, \textbf{RQ$_2$}, and \textbf{RQ$_3$}), we use multivariate linear regression to build seven models, one for each dependent variable in \tabref{tab:variables}. For example, to answer RQ$_0$, three regression models have been built, each using one of the three dependent variables relevant for this RQ.} 

\REVISED{The independent variable is always \emph{treatment}, while in terms of control variables we use all the ones in \tabref{tab:variables} but the \emph{Issue type}. Indeed, the dependent variables used in RQ$_0$, RQ$_2$, and RQ$_3$, differently from the one used in RQ$_1$, are meaningful only when applied to a whole code review (\eg the time spent to complete the code review, the confidence of the participant when submitting the review). Since a single review concerns several issues usually having a different type, for these RQs we do not consider the \emph{Issue type} as control variable.} 

\section{Results and Discussion} \label{sec:results}


\subsection{RQ$_0$: Differences in Reviews Output of Different Treatments}

\begin{figure*}
  \centering
  \includegraphics[width=0.53\linewidth]{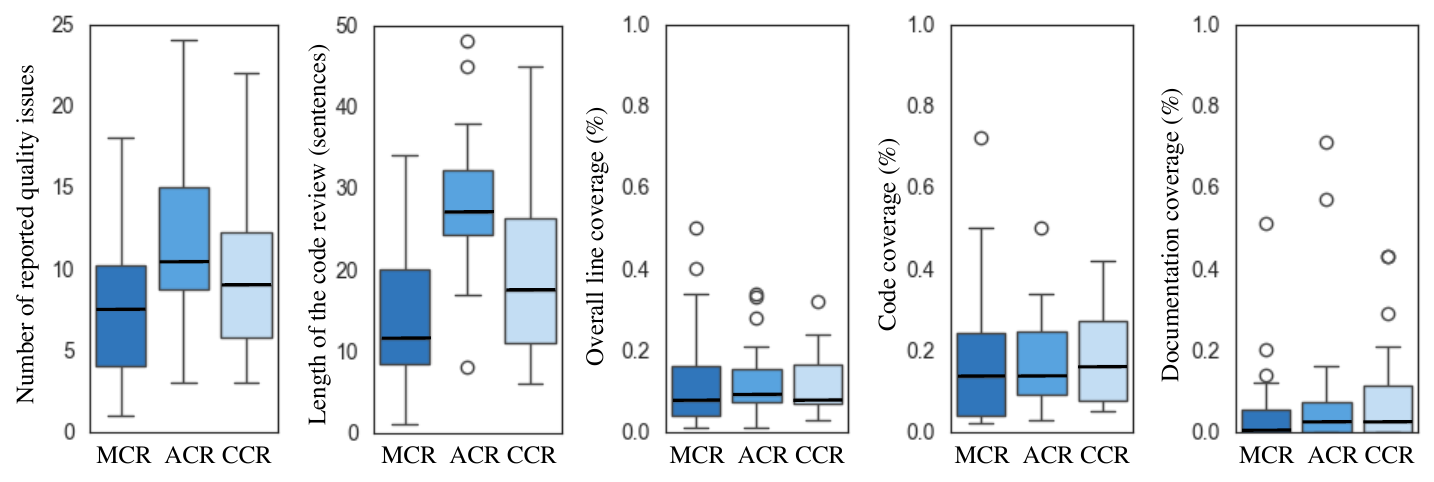}
  \vspace{-0.25cm}
  \caption{RQ$_0$: Characteristics of the final code reviews under the three treatments.}
  \vspace{-0.2cm}
  \label{fig:rq0_boxplot}
  
\end{figure*}

\figref{fig:rq0_boxplot} shows boxplots comparing the final version of the code reviews submitted by developers under the three treatments considered in our study, \ie \REVISED{\emph{manual code review} (MCR)}, \emph{automated code review} (ACR), and \emph{comprehensive code review} (CCR). From left to right we report boxplots comparing the number of quality issues reported in the review, the length of the review in terms of number of sentences, the overall line coverage of the review over the entire program (code and comments), as well as its coverage when considering only the source code, and only the documentation (comments). 

The two leftmost boxplots show clear differences among treatments in terms of the number of issues reported in the final reviews and in their length. Reviews resulting from treatments including an automated support (ACR and CCR) generally report more issues as compared to \REVISED{MCR}. \REVISED{The multivariate regression model (Multiple R$^2$ \cite{hughes1971statistics}: 0.32) --- \tabref{tab:linearRQ0} --- confirms the significant role played by the ACR treatment ($p$ $<$ 0.01) in the number of reported issues, with the Dunn's test \cite{dunn} showing a statistically significant difference when comparing ACR \emph{vs} \REVISED{MCR} ($p$-value = 0.0106 after Benjamini-Hochberg correction \cite{bh}).} \REVISED{Worth mentioning is also that participants tended to report more quality issues for some of the subject programs. This is expected considering that (i) we injected a different number of quality issues in each program, and (ii) ChatGPT identified a different number of quality issues in the programs, influencing the number of quality issues reported in the ACR treatment.}

\begin{table*}[h!]
	\centering
	\scriptsize
	\caption{\REVISED{RQ$_0$: Multivariate linear regression models (Estimate, Std. Error, Significance).}\vspace{-0.3cm}}
        \label{tab:linearRQ0}
	\begin{tabular}{l rrr c rrr c rrr}
		\toprule
		& \multicolumn{3}{c}{\emph{N. of reported quality iss.}} & & \multicolumn{3}{c}{\emph{Length of the code review}} & & \multicolumn{3}{c}{\emph{Covered code locations}}\\
		& \textbf{Estim.} & \textbf{S.E.} & \textbf{Sig.} && \textbf{Estim.} & \textbf{S.E.} & \textbf{Sig.} && \textbf{Estim.} & \textbf{S.E.} & \textbf{Sig.}\\\midrule
Intercept &  2.460 & 3.962 &    && 4.122 & 8.439& && -17.611 & 35.131\\
ACR &  4.269 & 1.331 & **  &&   11.862 & 2.836 & ***&& 5.297 &  11.806\\
CCR &  1.770 & 1.338 &   &&   2.549 & 2.850 & && -0.798 & 11.867\\
Years of experience & -0.032 & 0.071 &  &&  -0.098 & 0.151 & && 0.582 &  0.632\\
Involved in code review: Contributor \& Reviewer & 5.357 & 3.532 &   &&  11.386 & 7.522 & && 17.010 & 31.316\\
Involved in code review: None & 1.486 & 3.794 &   &&  5.473 & 8.081 & && 51.388 & 33.642\\
Involved in code review: Reviewer & 3.252 & 4.273 &   &&  -1.952 & 9.101 & && 28.911 & 37.886\\
Programming language: Python & -1.734 & 1.134 &  &&   -2.514 & 2.416 & && 3.215 & 10.058\\
Program: number-conversion &  0.437 & 1.649 &  &&   -0.344 &  3.513  & && 18.580 & 14.626\\
Program: stopwatch &  3.648 & 1.811 & *    &&  13.404 & 3.859  & ***  && 30.209 & 16.064\\
Program: tic-tac-toe &  5.426 & 2.106 & *  &&  13.041 & 4.486 &  **  && 11.427 & 18.676\\
Program: todo-list &  3.448 & 2.230 &   &&  5.860 & 4.750 &   && 13.780 & 19.773\\
Program: word-utils & 0.554 & 1.809 &   &&   2.591 & 3.854 &    && 21.437 & 16.046\\				
		\bottomrule
		\multicolumn{12}{c}{Sig. codes: `***' $p$ $<$ 0.001, `**' $p$ $<$ 0.01, `*' $p$ $<$ 0.05}
	\end{tabular}
	\vspace{-0.3cm}
\end{table*}

The final reviews of the ACR treatment identified, on average, 11.8 issues (median=10.5), against the 9.6 of CCR (median=9) and the 7.7 of \REVISED{MCR} (median=7.5). Such a result could be explained in part by the fact that reviewers kept most of the issues already present in the \emph{automated} (ACR) and in the \emph{comprehensive} (CCR) code review. 

Indeed, the ACR and CCR reviews we provided to participants featured on average 8.8 (median=9) and 4 (median=4) issues reported, respectively, and reviewers kept on average 7.1 (median=7.5) and 4.0 (median=4) of these issues, respectively. This leads to a first  outcome of our study: \faLightbulbO~Reviewers considered as valid most of the issues identified by ChatGPT Plus (by keeping them in their final review).  

\begin{figure}[h!]
\vspace{-0.2cm}
    \centering
    \includegraphics[width=0.4\linewidth]{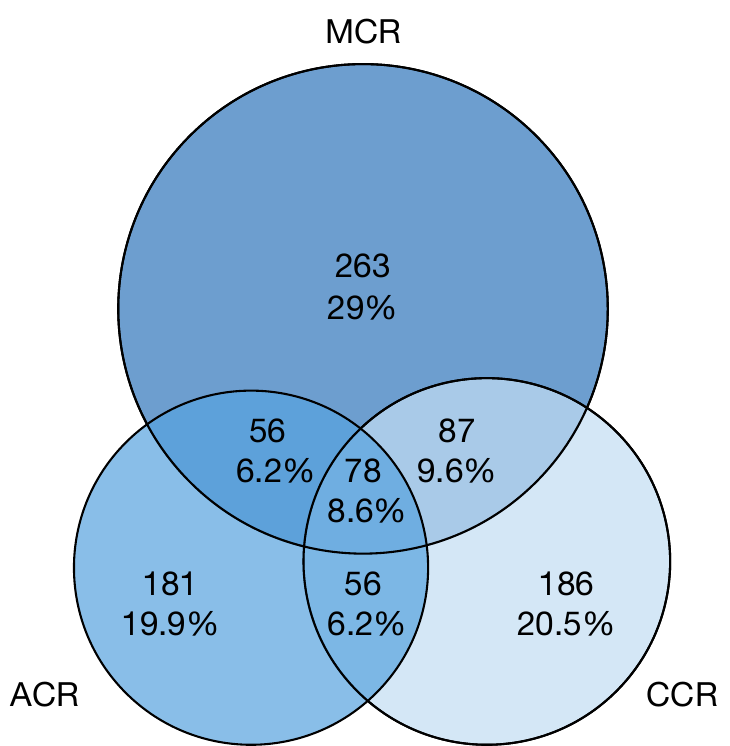} 
    \caption{RQ$_0$: Number of distinct lines covered (\ie commented on by participants) in the final reviews of the three treatments and their overlap.}
    \label{fig:rq0_venn}
\end{figure}

The fact that ACR and CCR reviews contained issues that reviewers kept in most cases had an impact on the final length of the submitted reviews, with ACR and CCR reviews being longer than those in the \REVISED{MCR} treatment (see \figref{fig:rq0_boxplot}). 

\REVISED{The regression model in \tabref{tab:linearRQ0} (Multiple R$^2$:  0.44) reports a significant impact of the ACR treatment ($p$ $<$ 0.001), with the Dunn's test \cite{dunn} confirming the statistically significant difference in length when comparing reviews output of the ACR and CCR treatments ($p$-value = 0.0046) and those output of ACR and \REVISED{MCR} ($p$-value = 0.0001) --- $p$-values adjusted with Benjamini-Hochberg correction \cite{bh}.}

To get a better understanding of the magnitude of such differences, the final reviews in the \REVISED{MCR} treatment include, on average, 16.3 sentences (median=11.5) compared to the 27.79 (median=27) in the final reviews of the ACR treatment. Interestingly, as our results show, a more verbose review does not necessarily mean a review that covers more code locations. The three rightmost boxplots in \figref{fig:rq0_boxplot} illustrate this phenomenon: there is no clear difference in the coverage (overall, on code, and on comments) between the final reviews of the three treatments (as also confirmed by the regression model in \tabref{tab:linearRQ0}). Moreover, the reviews written by participants without automated support had a higher variability in terms of the lines commented on by the different reviewers, while those starting from automated reviews tended to stay focused on the lines of code highlighted in the provided reviews. This is illustrated in the Venn diagram in \figref{fig:rq0_venn}, which shows the total number of different lines covered by all reviews belonging to each treatment. As observed, \REVISED{MCR} reviews covered a total of 484 distinct lines and 263 of these lines were unique to \REVISED{MCR}, \ie not covered by the ACR nor CCR final reviews. This was followed by CCR reviews, which covered 407 lines, 186 of which were unique to final reviews within this treatment, and lastly ACR reviews, covering 371 distinct lines, 181 of which being found only in ACR reviews.

\begin{figure}[t]
    \centering
    \includegraphics[width=0.9\columnwidth]{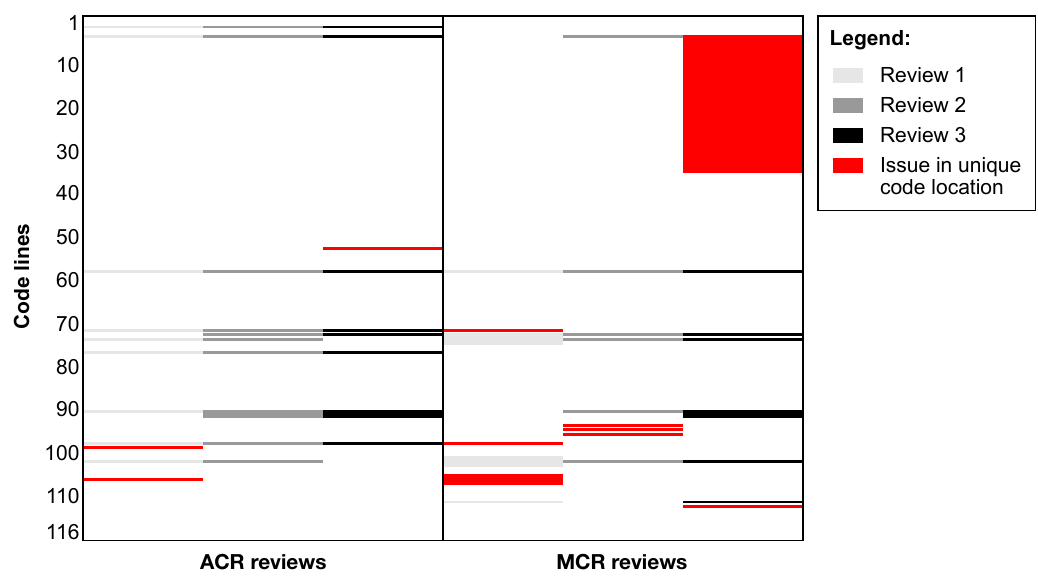}
    \caption{Example of distinct code lines covered by different reviews.}
    \label{fig:rq0_lines}
    \vspace{-0.3cm}
\end{figure}

\figref{fig:rq0_lines} shows an example of the code lines covered by three participant reviews (clear gray, gray, black) from the ACR (left) and \REVISED{MCR} (right) treatments for the \program{number-conversion} program in Java. Each rectangle denotes a single issue identified in the reviews, possibly spanning multiple lines (thicker rectangles). Red rectangles denote issues covering unique code locations (among the three reviews of that treatment). While for the ACR treatment there were only three issues from two reviewers covering unique locations, for the \REVISED{MCR} treatment, all three reviewers found a total of eight issues covering unique code lines, not covered in the other reviews. These findings result in two additional takeaways of our study. 

\eject

First, \faLightbulbO~reviews obtained with the support of automated tools might be more expensive to process for the contributor (\ie the developer submitting the code for review), since they are significantly more verbose as opposed to those manually written, while commenting on a similar amount of code lines. 

\eject

This might indicate an additional cost on the contributor's side, which complements the analysis we will present in RQ$_2$ when assessing the time spent by reviewers under the three treatments. Second, \faLightbulbO~the availability of an automated review influences the reviewer's behavior, who will mostly focus on the code locations commented on in the provided review. This also results in a lower variability in the types of issues identified by  reviewers for the same program. 


\begin{figure}[t]
	\centering
	\includegraphics[width=0.47\linewidth]{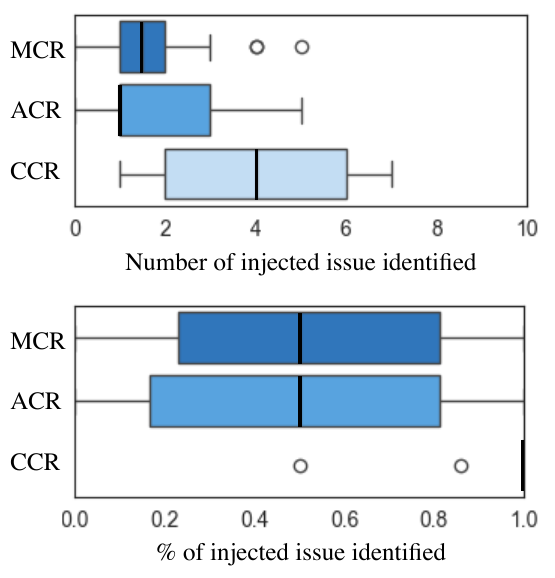}
	\vspace{-0.2cm}
	\caption{RQ$_1$: Number and percentage of identified injected issues.}
	\label{fig:rq1_boxplot}
	\vspace{-0.4cm}
\end{figure}

\begin{table}[h!]
	\centering
	\scriptsize
	\caption{\REVISED{RQ$_1$: Logistic regression model.}\vspace{-0.3cm}}
        \label{tab:logistic}
	\begin{tabular}{l rrr}
		\toprule
		& \textbf{Estim.} & \textbf{S.E.} & \textbf{Sig.}\\
		\midrule
Intercept & 1.371 & 1.581 & \\
ACR & -0.004 & 0.351  \\
CCR & 4.710 & 0.792 & ***\\
Years of experience &  -0.017 & 0.022  \\
Involved in code review: Contributor \& Reviewer & 0.308 & 0.906 & \\
Involved in code review: None & 0.143 & 0.850 & \\
Involved in code review: Reviewer & 0.437 & 0.977 \\ 
Programming language: Python & -0.589 & 0.403 \\
Program: number-conversion & 0.788 & 0.768  \\
Program: stopwatch & -1.380 & 0.727 \\
Program: tic-tac-toe & -0.084 & 0.865 \\
Program: todo-list & 1.298 & 1.049  \\
Program: word-utils & -2.029 & 0.706 & ** \\
Issue Type: Evolv. $\rightarrow$ Docum. $\rightarrow$ Textual &-1.032 & 1.088 \\
Issue Type: Evolv. $\rightarrow$ Structure $\rightarrow$ Org. &  -0.814 & 1.198 \\
Issue Type: Evolv. $\rightarrow$ Structure $\rightarrow$ Solution App. &0.014 & 1.112  \\
Issue Type: Funct. $\rightarrow$ Check & -0.649 & 1.144 \\
Issue Type: Funct. $\rightarrow$ Interface  &-1.546 & 1.263 \\
Issue Type: Funct. $\rightarrow$ Logic  &-1.838 & 1.583 \\
				
		\bottomrule
		\multicolumn{4}{c}{Sig. codes: `***' $p$ $<$ 0.001, `**' $p$ $<$ 0.01, `*' $p$ $<$ 0.05}
	\end{tabular}
	\vspace{-0.6cm}
\end{table}

\subsection{RQ$_1$: Impact on Quality Issues Found}

\figref{fig:rq1_boxplot} shows boxplots with the number (top) and percentage (bottom) of injected issues identified in the final reviews of the three treatments. \REVISED{\tabref{tab:logistic} reports the results of the logistic regression model using the \emph{Is injected bug identified} as dependent variable}. As expected, the final reviews submitted  under the CCR treatment usually report 100\% of injected issues, as these were already present in the review initially provided to them and kept in the final review (with the logistic regression confirming the significant influence of CCR on the odds of identifying the injected bug). 

More interestingly, the final ACR reviews submitted by participants uncovered a median of 50\% of the injected issues, \ie the same amount uncovered in the \REVISED{MCR} reviews, which were written from scratch by participants, despite the fact that the automated reviews initially provided in the ACR treatment already reported, on average, 42\% of the injected issues. This unveils a few important findings of our study: \faLightbulbO~the current state of the art in terms of automated code review does not lead to a higher number of injected issues being identified compared to fully manual code reviews. This is due to two factors. First, ChatGPT produced reviews in which more than half of the injected issues, on average, were not identified. Second, the exposure that developers had to these automated reviews before writing their final review seems to have biased their behavior, leading to them barely identifying any additional injected issues compared to the ones already included in the automated review. These findings should act as a clear warning sign for companies interested in adopting automated support in code review. A possibility to consider is to provide the automated review only once the reviewer already submits their comments, thus not being biased by the ones already provided.



Since the final reviews of ACR report the highest number of issues without, however, identifying a higher number of injected issues, a question arises about the relevance of the additional issues identified in the ACR reviews. Indeed, it is possible that additional issues reported are just as relevant as the injected ones. To assess this, we asked two developers not involved in the study to assess the severity of the: (i) injected issues as documented in the reviews provided as starting point in the CCR treatment; (ii)~non injected issues automatically identified in ACR reviews; and (iii)~non injected issues manually identified in \REVISED{MCR} reviews. The two developers have 16 and 7 years of programming experience, respectively, and were instructed to provide a severity assessment on a scale from 1=\emph{low severity} to 3=\emph{high severity}, with the former indicating issues which they do not consider mandatory to address for approving the code and the latter indicating showstoppers. 

While we acknowledge the subjectivity of this assessment, the two developers who performed it had a strong disagreement (\ie 1 \emph{vs} 3) in only 5\% of the inspected issues. 

The weighted $k$ agreement \cite{cohen:lawrence1988} was 0.315. The findings of this analysis indicated that the issues we injected were the ones assessed with the highest severity (Q1=2.0, Q2=2.0, Q3=3.0, mean=2.2), followed by the additional ones manually identified by developers (Q1=1.0, Q2=2.0, Q3=2.0, mean=1.8), and the additional ones recommended by ChatGPT (Q1=1.0, Q2=2.0, Q3=2.0, mean=1.6). \REVISED{We also factored in the issue severity as a further cofactor in the logistic model without, however, observing any impact of it nor changes in the significant variables (output of the model available in \cite{replication}).}

The difference between the severity of the injected issues and the additional ones identified either manually or automatically is statistically significant ($p$-value $<$ 0.01) with a medium effect size (Mann-Whitney test \cite{wilcoxon} and the Cliff's delta \cite{Cliff:2005}). An example of an automatically identified issue classified as low severity by both developers is: ``\emph{This method provides an interesting feature by $[\dots]$. The logic is sound, though it involves several conditionals and might benefit from comments explaining the rationale for each case.}''. In this case, while comments are indeed missing, the code is quite self-explanatory. On the other hand, some high-severity injected issues were missed by ChatGPT, such as the following one, provided in the initial review of the CCR treatment: ``\emph{There is a critical bug. For HEXADECIMAL, it erroneously uses base 8 instead of 16. It should be decimalToAnyBase(num, 16).}''.

RQ$_1$'s findings lead to the conclusion that the reviews output of the ACR treatment identify a higher number of low-severity issues while, however, not making a difference when it comes to spotting high-severity issues (\ie the ones we injected) as compared to a full manual process. This supports our former suggestion that, \faLightbulbO~given the current state of automation, automatically generated reviews may be considered as a useful complement at the end of a manual review process.

\subsection{RQ$_2$: Impact on Review Time}

\begin{figure}[t]
\centering
	\includegraphics[width=0.85\linewidth]{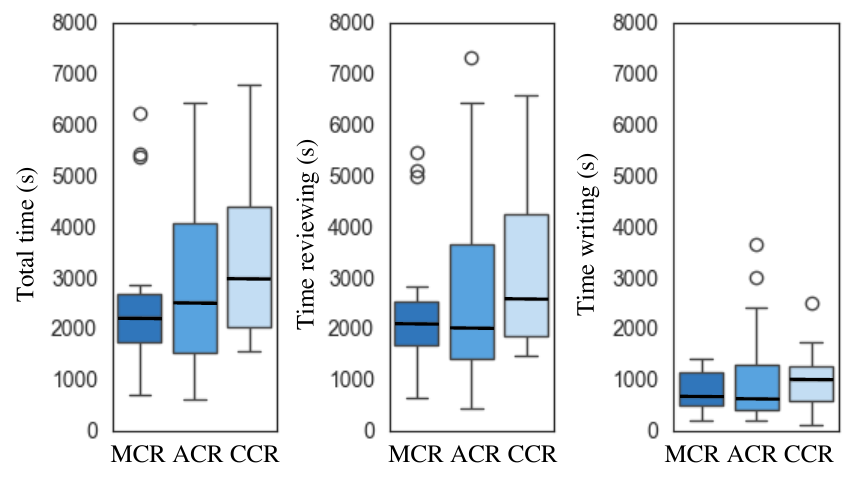}
	\caption{RQ$_2$: Time spent in reviews across different treatments.}
	\label{fig:rq2_boxplot}
	\vspace{-0.3cm}
\end{figure}

\figref{fig:rq2_boxplot} shows from left to right: the time (in seconds) that reviewers spent on the whole code review process, the time spent on the code to review (\ie not including running the program or the tests), and the time writing the review comments. For the ACR and CCR treatments, the latter also includes the time spent \emph{reading} the originally provided reviews. \REVISED{While we also created multivariate linear regression models, we found that none of the involved independent variables (treatments and cofactors) plays a statistically significant role on any of the time-based dependent variables. Thus, for the sake of space, we only report these models in our replication package \cite{replication}.}
\faLightbulbO~This finding debunks one of the motivations for automated code review \cite{tufano:icse2022}, \ie saving time for reviewers. In fact, \figref{fig:rq2_boxplot} shows that reviews performed completely manually (\REVISED{MCR}) took less time (mean=42, median=37 minutes) than those supported by automation in the ACR (mean=56, median=42 minutes) and CCR (mean=57, median=50 minutes) treatments. 

This might be due to the fact that the introduction of an \emph{automated} review, even if correct (CCR), comes with a price, namely the reading, understanding, and double-checking of the provided comments. This seems to be the case especially for CCR reviews, where reviewers spent an average of 17 minutes (median=16) reading and writing reviews, compared to an average of 12 minutes (median=11) for reviews written from scratch (see rightmost boxplot in \figref{fig:rq2_boxplot}).

It is also interesting to note that the variability in the time spent on the code to review and on the overall review process is higher for the ACR and the CCR treatments as compared to \REVISED{MCR}. We presume that this is due to two factors, namely: (i)~the length of the automatically generated reviews, where the shortest pointed to no issues (ACR review of \program{stopwatch} in Python) while the longest pointed to 14 issues (ACR review of \program{maze-generator} in Java); and (ii) the trust that reviewers put in the automated support, which may vary from one reviewer to another. \faLightbulbO~Longer case studies in which developers have time to build their opinion about the review automation tool are needed to corroborate or contradict our findings, especially when it comes to what we observed in terms of time spent.

\begin{table}[h!]
\vspace{-0.2cm}
	\centering
	\scriptsize
	\caption{\REVISED{RQ$_3$: Linear regression model.}\vspace{-0.3cm}}
        \label{tab:linearRQ3}
	\begin{tabular}{l rrr}
		\toprule
		& \textbf{Estim.} & \textbf{S.E.} & \textbf{Sig.}\\
		\midrule
Intercept &   2.148 & 0.628 & ** \\
ACR & 0.099 & 0.211    \\
CCR & 0.180 & 0.212    \\
Years of experience &  0.025 & 0.011 & *  \\
Involved in code review: Contributor \& Reviewer & 1.001 & 0.560    \\
Involved in code review: None & 1.147 & 0.602    \\
Involved in code review: Reviewer & 1.591 & 0.678 &  *  \\
Programming language: Python & 0.270 & 0.180    \\
Program: number-conversion &  0.361 & 0.261    \\
Program: stopwatch & -0.296 & 0.287   \\
Program: tic-tac-toe & 0.233 & 0.334    \\
Program: todo-list & -0.040 & 0.353   \\
Program: word-utils & -0.592 & 0.287 & *\\		
		\bottomrule
		\multicolumn{4}{c}{Sig. codes: `***' $p$ $<$ 0.001, `**' $p$ $<$ 0.01, `*' $p$ $<$ 0.05}
	\end{tabular}
	\vspace{-0.3cm}
\end{table}

\subsection{RQ$_3$: Impact on Reviewer's Confidence}

After each code review task, reviewers scored their confidence in the review they submitted on a scale from 1 (\emph{very low confidence}) to 5 (\emph{very high confidence}). Reviews from the \REVISED{MCR} treatment were scored with an average confidence of 3.5 (median=4), those from the ACR treatment with an average confidence of 3.7 (median=4), and those from the CCR treatment with an average confidence of 3.8 (median=4). \REVISED{Indeed, as shown in the linear regression model in \tabref{tab:linearRQ3} (Multiple R$^2$: 0.3445), there is no significant impact of the treatment on the confidence score reported by reviewers}. We thus conclude that \faLightbulbO~providing an automated code review as a starting point, even one being able to identify several high-severity issues (\ie CCR treatment), does not have a significant effect on the confidence of the reviewer. This might be due to the fact that, while the automated code review may point the reviewer to relevant code locations, it does not help in understanding the code which, in the end, is what we expect to mostly influence a reviewer's confidence. Combining automated code review with LLM-based code summarization \cite{Toufique:ase2022} may help in that direction.


\REVISED{\subsection{Actionable Recommendations}}

\REVISED{Based on our findings, we distill the following actionable recommendations for reviewers, designers of tools aimed at automatically generating code reviews, and researchers.}

\REVISED{
\emph{Reviewers:} We observed that the availability of an automated review strongly influences the reviewer's behavior, who will mostly focus on the code locations commented on in the provided review. Also, automated reviews result in a lower variability in the types of issues identified by different reviewers for the same program. Based on these findings, we recommend to adopt automated reviews as a further check only after the manual inspection. This will not save time but it could help in identifying additional quality issues.}

\REVISED{
\emph{Tools' designers:} While most of the issues reported by ChatGPT have been considered valid by reviewers (\ie kept in the final review), we found that these issues tend to have a quite low severity. Tools tailored for the identification of specific high-severity quality issues would be a valuable asset. Also, automatically generated reviews are much more verbose than those manually written. This is a non-negligible cost that should be considered in the design of tools, \eg by making a best effort to keep the generated reviews concise.}

\REVISED{
\emph{Researchers:} We did not observe any time saved thanks to the availability of automated reviews. Thus, the motivation for introducing these tools in a code review process may be more related to a more comprehensive code inspection rather than to save time \cite{tufano:icse2022}. Also, given the strong bias in reviewers' behavior we identified, studies investigating the impact on practitioners' behavior when exploiting AI-based tools to (semi-)automate SE tasks are very much needed.
}
\section{Threats to Validity} \label{sec:threats}

\textbf{Construct validity.} A major challenge was the time measurement in RQ$_2$, since interruptions were possible while the participants were performing the code reviews. We instructed participants to not interrupt a code review task and to take breaks only when changing treatment. Still, in the time-based analysis we excluded 13 data points (out of 72) since it was clear that they represented errors in the measurement when compared to the other timings. The removed data points were balanced across the treatments (4 for NCR, 5 for ACR, and 4 for CCR) and did not change the final outcome of our study. 

Analyses reported in RQ$_0$ and RQ$_1$ are based on data manually extracted from the submitted reviews (\eg number of issues reported in the reviews), thus involving subjectivity risks. 
To partially address them, we made sure that each review was independently inspected by two evaluators.

\textbf{Internal validity.} Participants who only contributed one or two of the assigned reviews (18 out of \participantsSelected contributed three reviews) made the study diverge from our initially planned \emph{within-subject} design. Still, we addressed the issue by only considering \selectedReviews of the reviews we collected in our analysis, with the goal of balancing within each treatment the number of reviews performed on each project.  

\REVISED{While we collected some demographic data about participants, there are several other factors that could have influenced our findings. These include: (i) educational background; (ii) work experience; (iii) the lack of knowledge some participants may have about the inspected programs; (iv) no past experience of participants in using code review automation tools, thus not being able to properly calibrate their confidence in accepting/rejecting the recommendations; and (v) possible time constraints that participants had while performing the review task. Our replication package \cite{replication} features a causal diagram showing all measured and unmeasured variables which may have played a role on the measured dependent variables.}

\REVISED{\textbf{External validity.} As a design choice, we decided to only involve in our study professional developers. This resulted in a limited number of participants who took part to our study (29) for a total of 24 data points (\ie submitted review) per treatment. We acknowledge that our study could be statistically underpowered, thus leading to biased conclusions. For this reason, replications are needed in order to corroborate or contradict our preliminary observations.}

\REVISED{Another threat to the generalizability of our findings concerns the representativeness of the issues we injected as representative of those actually found in industrial code review. For example, concerns may arise about the triviality of the injected issues, also considering the subject programs which, for the sake of limiting the study time, were limited in terms of size. Nonetheless, we took inspiration from the taxonomy of issues found in code reviews documented by M\"antyl\"a and Lassenius~\cite{mantyla2008types}, trying to inject a proportion of evolvability and functional issues close to the one they document (\ie $\sim$3 out of 4 issues found in code review do not pertain the visible functionality of the program). Also, our results show that, when not running the task in the context of the CCR treatment (\ie a review identifying the injected issues was available), participants were able to identify all injected issues in only 12 out of the 48 code review tasks. This addresses concerns about the triviality of at least identifying the issues.}
\section{Related Work} \label{sec:related}


\subsection{Code Review Automation}

Most of the prior work focused on classification tasks, such as recommending the best suited reviewer(s) for a given change \cite{balachandran:icse2013, thongtanunam:saner2015, xia:icsme2015, ouni:icsme2016, rahman:icse2016, zanjan:tse2016, asthana:esecfse2019, jiang:jss2019, mirsaeedi:icse2020, strand:icse-sep2020, pandya:esecfse2022}, classifying the sentiment of review comments \cite{ahmed:ase2017,egelman:icse2020}, and their usefulness \cite{pangsakulyanont:iwesep2014, rahman:msr2017, hasan:emse2021}, \etc 

More relevant to our work are the generative code review tasks recently automated via DL. The two most commonly addressed tasks in the literature are \emph{review comment generation} \cite{li.l:esecfse2022, li:fse2022, hong:esecfse2022, tufano:icse2022} and \emph{code refinement} \cite{huq:ist2022, tufano:icse2021, tufano:icse2022, li:fse2022}. 
The former consists in automatically generating review comments in natural language for a given piece of code, similar to those that a human reviewer would write. The input of the DL model is represented by the code to review, while its output consists of a set of natural language comments pointing to issues in the code. In the code refinement task, the DL model takes as input the code submitted for review and the natural language comments written by a human reviewer. The goal of the model is to \emph{refine} the code to address the reviewer's comments. The evaluations of these tasks showed promising results (\eg $\sim$30\% correct predictions in the code refinement task \cite{li:fse2022}), while still pointing to the need for major improvements before these tools can be considered for industrial adoption \cite{Tufano:tse2024}. 

Given the documented evidence of ChatGPT's adoption in open source projects as a co-reviewer \cite{Tufano:msr2024}, in our experiment we adopted ChatGPT as the representative for code review automation tools supporting the \emph{comment generation} task.

\begin{table*}[ht]
	\centering
    \caption{Controlled experiments on code review: For participants' type, ``S'' indicates students, ``P'', practitioners.\vspace{-0.3cm}}
    \label{tab:experiments}
    \resizebox{\textwidth}{!}{%
    {\scriptsize
    \renewcommand{\arraystretch}{0.6}
    \begin{tabular}{@{}lrclrp{5.1cm}p{4.8cm}@{}}
    \toprule\vspace{0.05cm}
    {\bf Reference} & \multicolumn{2}{c}{{\bf Participants}} & {\bf Programming} & {\bf \#Review} & {\bf Independent Variable} & {\bf Main Dependent}\\\cline{2-3}
    & {\bf \#} & {\bf Type} & {\bf Languages} & {\bf Tasks} & {\bf Manipulated} & {\bf Variables Measured}\\\midrule
    
    Runeson and Wohlin \cite{Runeson:emse1998} & 8 & P, S & C & 24 & Three approaches for estimating the number of bugs in the code being reviewed & How close the approaches were in terms of estimates after the review was performed\\\midrule 
    
    Tao and Kim \cite{Tao:msr2015} & 18 & S & Java & 36 & Using/not using an approach to automatically partition composite  changes submitted for review & Code review correctness, Review time\\\midrule 
    
    Zhang \etal \cite{Zhang:icse2015} & 12 & S & Java & 24 & Using/not using an interactive approach to inspect code changes &  Comprehension level of reviewed change assessed via a questionnaire\\\midrule 
    
    Khandelwal \etal \cite{Khandelwal:isec2017} & 183 & S & Python & 915 & Using a gamified/non-gamified code review tool & Usefulness of code review comments, Identified bugs/code smells \\\midrule
    
    Huang \etal \cite{Huang:ase2018} & 10 & S & Java & 10 & Using/not using a code differencing tool to simplify the understanding of code changes & Comprehension level of reviewed change assessed via a questionnaire\\\midrule

	Huang \etal \cite{Huang:fse2018} & 14 & P, S & Java & 28 & Providing/Not-providing reviewers with information about the silent class in a commit (\ie the one triggering changes to other modified classes) & Comprehension level of reviewed change assessed via a questionnaire\\\midrule
	
    Hanam \etal \cite{Hanam:icsme2019} & 11 & P, S & Javascript & 22 & Using/not using a tool performing change impact analysis & Identified bugs, Review time\\\midrule
    
    
    Spadini \etal \cite{Spadini:icse2020} & 85 & P, S & Java & 85 & Showing/not showing already existing review comments to a reviewer starting their code inspection  & Identified bugs\\\midrule 
    
    Fregnan \etal \cite{Fregnan:fse2022} & 106 & P, S & Java & 106 & Manipulating the inspection order of files impacted by a change &  Identified bugs in files appearing in different positions\\\midrule 
    
    {\bf Our study} & {\bf 29} & {\bf P} & {\bf Java, Python} & {\bf 72} & {\bf Three treatments providing/not providing reviewers an automatically generated review} & {\bf Identified bugs, Review time, Reviewer's confidence}\\
    
    \bottomrule

    \end{tabular}
    }
    }
    \vspace{-0.3cm}
\end{table*}

\subsection{Controlled Experiments on Code Review}

There are several prior controlled experiments on code review reported in the literature.  \tabref{tab:experiments} offers a brief overview of these works and reports (i) the related reference, (ii) the number and type of participants involved in the study (where type can be P = practitioners, S = students, or both), (iii)~the subject programming languages, (iv) the number of total review tasks collected (in some studies participants performed more than one task, while in others only one), and (v) a short description of the manipulated independent variables and measured dependent variables. The last row of \tabref{tab:experiments} reports the same information for our study.
As illustrated, all works focus on independent variables different from the one tackled in our study (\ie the presence/absence of an automated code review). Several works \cite{Zhang:icse2015,Huang:ase2018,Huang:fse2018} also differ in terms of measured dependent variables, focusing on the reviewer's comprehension level. We discuss in the following only the most related controlled experiments, being the ones sharing with us the measured dependent variables.

Khandelwal \etal \cite{Khandelwal:isec2017} study how the usage of gamified code review tools can improve the usefulness of the code review comments and the identified quality issues. The study involved 183 undergraduate students who had to write (i) a program which, on average, was composed by $\sim$500 lines of code, and (ii) reviews for 5 peer-written programs using one of five code review tools (three being gamified and two not). This led to a total of 183 $\times$ 5 code review tasks, which showed no impact of gamification on the code review quality.

Hanam \etal \cite{Hanam:icsme2019} present an experiment involving 11 practitioners and students to investigate the impact of change impact analysis on the ability of identifying bugs during code review. 
Each participant performed two code review tasks, one using a change impact analysis tool named SemCIA and one not using it, leading to 22 code reviews collected. 

Using SemCIA participants were able to perform the code review quickly while also identifying more bugs.

Spadini \etal \cite{Spadini:icse2020} study how showing already existing review comments to a reviewer starting their code inspection influences the number of identified bugs. This experiment involved 85 participants, including practitioners (57) and students. The authors asked each participant to perform one code review, for a total of 85 collected reviews. The authors show that existing review comments can help in identifying specific types of bugs that otherwise would have been overlooked. 

Finally, Fregnan \etal \cite{Fregnan:fse2022} investigate whether the order in which files submitted for review are presented to the reviewer impacts the number of bugs identified in them. Surprisingly, they found that by just changing the file position the odds of identifying bugs in the inspected files can substantially change, with those inspected first having higher odds. The study has been conducted with 106 among practitioners (72) and students. Also in this case, each participant contributed to only one treatment, with a total of 106 reviews collected.

As highlighted in \tabref{tab:experiments} and previously mentioned, the novelty of our work as compared to the existing literature lies in the investigated independent variable focusing on the novel generation of DL-based code review automation tools.

\section{Conclusion and Future Work} \label{sec:conclusion}

In this study, we explored the effects of incorporating automatically generated code reviews, particularly those produced by ChatGPT Plus (GPT-4), into the code review process. 

Our controlled experiment involved \participantsSelected professional developers who reviewed code in three settings: manually, having an automated review provided by ChatGPT Plus as a starting point, or starting from a review that was manually crafted by the authors to capture all major issues in the code, but rephrased by ChatGPT Plus to seem generated by it. 

Reviewers generally accepted the validity of issues identified by the LLM, adopting 89\% of them on average in their final reviews. However, the presence of an automated review influenced reviewers to concentrate on the highlighted code locations, potentially overlooking other areas. Our findings also indicate that using automated ChatGPT Plus reviews as a starting point leads to less severe issues being identified compared to manual reviews, but unveils more trivial issues. Additionally, while automated reviews are expected to save time, our findings show that, in practice, reviewers still had to spend time verifying the accuracy of the automated comments, negating any potential time savings. At the same time, having access to automated reviews as a starting point did not lead to changes in the confidence of reviewers. 

These findings suggest that while current LLMs can play a valuable role in identifying part of the code issues, their use as automated co-reviewers does not necessarily improve the efficiency or effectiveness of the code review process from a holistic perspective. The tendency of automated reviews to focus reviewer attention on specific areas of the code, coupled with their limited impact on identifying high-severity issues and inefficiency in reducing review time represent important challenges that should be addressed in future work to allow LLMs to become integral parts of the code review process.

\section*{Acknowledgment}
This project has received funding from the European Research Council (ERC) under the European Union's Horizon 2020 research and innovation programme (grant agreement No. 851720) and from the Swiss National Science Foundation (SNSF) under the project ``PARSED'' (grant agreement No. 219294). We are deeply grateful to the participants who took part in the study for their time and effort.

\eject

\bibliography{main}
\bibliographystyle{IEEEtran}

\end{document}